# Spatially resolved TiO$_x$ phases in RRAM conductive nanofilaments using soft X-ray spectromicroscopy

Daniela Carta, Adam Hitchcock, Peter Guttmann, Anna Regoutz, Ali Khiat, Alexantru Serb, Isha Gupta, Themistoklis Prodromakis

**Reduction in metal-oxide thin films has been suggested as the key mechanism responsible for forming conductive nanofilaments within solid-state memory devices, enabling their resistive switching capacity. The quantitative spatial identification of such filaments is a daunting task, particularly for metal-oxides capable of exhibiting multiple phases as in the case of TiO$_x$. Here, we spatially resolve and chemically characterize distinct TiOx phases in localized regions of a TiO$_x$–based memristive device by combining full-field transmission X-ray microscopy with soft X-ray spectroscopic analysis that is performed on lamella samples. We particularly show that electrically pre-switched devices in low-resistive states comprise reduced disordered phases with O/Ti ratios close to Ti$_2$O$_3$ stoichiometry that aggregate in a ~ 100 nm filamentary region electrically conducting the top and bottom electrodes of the devices. We have also identified crystalline rutile and orthorhombic-like TiO$_2$ phases in the region adjacent to the filament, suggesting that the temperature increases locally up to 1000 K, validating the role of Joule heating in resistive switching. Contrary to previous studies, our approach enables to simultaneously investigate morphological and chemical changes in a quantitative manner without incurring difficulties imposed by interpretation of electron diffraction patterns acquired via conventional electron microscopy techniques.**

Nanoscale resistive random-access memory devices (RRAM) based on metal-oxides, also known as memristors, have great prospect in becoming a mainstream memory technology, due to their infinitesimal dimensions, fast switching and capacity to store multiple bits of information per element. [1,2,3,4,5,6] Such devices have the ability to toggle their resistance between high resistive state (HRS, or ON) and low resistive state (LRS, or OFF) in either a digital or analogue fashion.[7] TaO$_x$[8] and HfO$_x$[9] are of particular interest for RRAM due to their high-$k$ properties and compatibility with deep submicron complementary metal-oxide semiconductors (CMOS) technologies and their simple reduction dynamics; the ability to toggle between two phases only, make them easy to control while at the same time enhance their state stability. On the other hand, TiO$_x$, one of the most celebrated metal oxides, provides wider opportunities for multi-state memory capacity due to the intrinsic variety of possible chemical phases. This complexity however makes its control more difficult, also prohibiting the rigorous study of the unrelying physical mechanisms. Whilst the electrical behaviour of TiO$_x$ based memristors has been widely investigated, much less is known about the atomic-scale changes that occur in the film as a consequence of switching.[10] It has been suggested that the formation and migration of defects within the oxide active layer induce the formation of reduced phases in form of conductive filaments that render resistive switching. However, direct localization and chemical characterization of such conductive phases remains a challenge.[11]

Previous work by Kwon et al.[12] based on high resolution transmission electron microscopy (HRTEM), identified Ti$_4$O$_7$ as the structure of conductive filaments in TiO$_2$ based-devices by interpretation of electron diffraction patterns. However, HRTEM requires the use of high energy electron beam (200-300 kV) which could induce sample crystallization and effectively alter the physical-chemical state of the film. Furthermore, electron diffraction requires the crystals to be aligned in specific orientations with respect to the beam for unambiguously capturing distinct diffraction patterns.[12] In this work, we employed soft X-ray spectromicroscopy for measuring Near-Edge X-ray Absorption Fine Structure (NEXAFS) spectra at high spatial resolution in a full-field Transmission X-ray Microscopy (TXM). This technique allowed us to perform simultaneously imaging and spectroscopy for investigating morphological changes in the film as well as performing chemical analysis at nanometric scale of localized regions, respectively. Our set-up differs substantially from previous transmission X-ray microscopy studies of $TiO_2$-based [13,14] or $SrTiO_3$-based [15] RRAM devices, which are based on irradiating the device from the top electrode, alleviating the challenges imposed for accessing the $TiO_x$ film through the capping top electrode. When compared to electron energy loss spectroscopy (EELS), TXM-NEXAFS provides more analytically useful information per unit radiation damage.[16] Moreover, identification of distinct phases using NEXAFS is based on fingerprints methods, avoiding the difficulties imposed via multiple scattering effects and/or the presence of diffraction spots from other $TiO_2$ phases and/or metallic electrodes when employing electron diffraction.

This technique has allowed us to spatially resolve distinct $TiO_x$ phases that macroscopically assemble in a conducting filament within a RRAM prototype that has been switched to a LRS. We prove that the conductive path is located right underneath a protrusion of the top electrode and it is composed by a reduced $TiO_x$ phase with a O/Ti ratio close to $Ti_2O_3$. Crystalline $TiO_2$ rutile phase and an additional orthorhombic-like $TiO_2$ phase regions were also identified in a localized area between the $TiO_x$ reduced phase and the top electrode, suggesting that the temperature due to Joule heating increases up to 1000 K in the core of the filament.[17] A Ti-containing phase similar to the $TiO_x$ reduced phase was also identified in both the top and bottom Pt electrodes, giving evidence of Ti diffusion into Pt due to the switching.[18]

The studied prototypes are Cr(3nm)/Pt(30nm)/$TiO_x$(50nm)/Pt(30nm) stacks (Fig. 1a), fabricated as stand-alone devices with a junction of 40x40 $\mu m^2$ (Fig. 1b and 1c) (see Methods). The as-deposited $TiO_x$ film was determined to be close to stoichiometry by X-ray photoelectron spectroscopy (Supplementary, Fig. S1). Prior the TXM-NEXAFS investigation, all devices were electrically characterized using voltage sweeping with their distinct memory electrical states kept as pristine (PRI) and set as LRS as respectively shown in the current-voltage (I-V) characteristics of Fig. 1d and 1e.

TXM-NEXAFS measurements were performed on a cross-section of each of the two devices cases, *i.e.* PRI and LRS. An outline of the method is briefly described in Fig. 2. The lamella is extracted from the red region highlighted in Fig. 2a using the Focused Ion Beam (FIB) lift-out technique (see Methods). The thin lamella (Fig. 2b) is positioned perpendicular to the X-ray beam direction for absorption measurements and imaging as shown in Fig. 2c. A schematic of the TXM experiment is shown in Fig. 2d. For each of the two cases, PRI and LRS, two sequences (stacks) of X-ray images were acquired at closely spaced photon energies using a transmission X-ray microscope, in the Ti 2$p$ (450-485 eV) and O 1$s$ (525-555 eV) energy range. Each stack, consisting of 1053 images over the Ti 2$p$ energy range and 351 images over the O 1$s$ energy range, was carefully aligned using a cross-correlation iteration process until the image shift was less than ±0.2 pixels (± 1 nm) across the entire energy range (see Methods). After conversion of all stacks into optical density (absorbance), a detailed analysis involving extraction of the X-ray absorption spectra of specific regions of interest and conversion to component maps using spectral fitting or multivariate statistical analysis procedures was performed.[19,20]

X-ray images of the lamellae for the PRI case at 450 eV (below the Ti 2$p$ edge) and 465 eV (on the strongest Ti 2$p$ absorption peak) are presented in Supplementary, Fig. S2. The individual layers comprising the device stack *i.e.* Si wafer, SiO$_2$, Pt bottom electrode (BE), TiO$_x$ active film and Pt top electrode (TE) can be clearly distinguished in Fig. S2a; the TiO$_x$ active layer appears dark as Ti does not absorb at this energy. At 465 eV, the TiO$_x$ layer appears bright due to the strong absorption of Ti (Fig. S2b). Spatially localized NEXAFS spectra at the Ti 2$p$ and at the O 1$s$ were then extracted from the TiO$_x$ film area. Two representative spectra are reported in Fig. S2c and S2d, respectively. The fine structure of Ti 2$p$ spectrum is complex but at a coarse level can be considered to consist of four main peaks (two doublets). The first doublet (2$p_{3/2}$) occurs within 457-462 eV while the second doublet (2$p_{1/2}$) appears in the range 462-468 eV (see Supplementary).[21] The Ti 2$p$ spectrum exhibits distinct features for the different polymorphs phase of crystalline TiO$_2$ and for amorphous TiO$_2$.[22] In particular, peaks of amorphous TiO$_2$ are quite broad due to structural disorder caused by a range of bond angles and lengths, in contrast to spectra of crystalline samples which are characterized by sharp peaks. The Ti 2$p$ spectrum (Fig. S2c) is consistent with amorphous TiO$_x$.[22,23] This result is in agreement with a TiO$_x$ thin film prepared via reactive sputtering and indicates damage/crystallisation was not caused by the sample preparation process. In agreement with our observations at the Ti 2$p$, the O 1$s$ spectrum (Fig. S2d), possesses typical signatures of disordered systems,[22] demonstrating that the TiO$_x$ film is amorphous all along the cross-section.

Notable differences are however observed in the film after switching to LRS (Supplementary, Fig. S3 and S4). Optical (Fig. S3b and S4a) and AFM images (Fig. S4b) viewed from the Pt TE show the formation of a protrusion about 140 nm high at the top left rim of the TE. Several studies have shown that protrusions of the TE could indicate possible critical regions of the film responsible for the RS switching.[24,25] The LRS lamella cut across the TE defect (Fig. S4c), presents some electron transparent areas in the damaged region (Fig. S4d) indicating the presence of a very low absorption area. Under the identified protrusion, the TE is bulged and discontinuous, having the shape of a dome (Fig. S4e). The region below this dome is formed by a main open cavity with two additional smaller ones towards the region outside the junction. The TiO$_x$ layer is dislocated towards the region where the TE is discontinuous. The dome-like empty area, suggests that the TE protrusion was caused by O$_2$ gas escaping through the weakest point of the TE, the rim, in agreement with previous observations.[24,25] These features are similar to those reported in ref [11].

The X-ray cross-section image of the LRS case recorded in the region underneath the TE protrusion at the energy of 450 eV is reported in Fig. 3a. The Pt/TiO$_x$/Pt stack can be clearly distinguished on the right side of Fig. 3a (Pt BE/TE are bright and TiO$_x$ is black) and appear unaffected. In the defect region, the BE remains unaffected whereas the TE is discontinuous and delaminated from the TiO$_x$ film. At 450 eV, the Ti containing regions cannot be distinguished from void zones as both appear dark. By examining the Ti 2$p$ image sequence at different energies, we can spatially resolve the location of regions containing Ti species as the contrast of the image changes in correspondence with changes in the Ti absorption peak intensities. The image at 458 eV (Fig. 3b) shows highly absorbing regions containing TiO$_x$ appearing with brighter contrast; the regions that appear dark at 450 eV but bright at 458 eV correspond to Ti containing areas. Regions that remain dark at 458 eV are clearly defined as voids. It is interesting to note that three highly localized (filamentary) Ti containing regions, (indicated as A, B and C – see Fig. 3b) link the BE and TE. In particular, the magnified image of Fig. 3c, shows that the TiO$_x$ layer aggregates in a ~ 100 nm wide localized region underneath the highest deformation of the TE (C). These observations are consistent with the X-ray coloured image shown in Supplementary, Fig. S5 where the areas containing only Ti species are shown in red. By extracting NEXAFS spectra in the regions where Ti species exist, chemical identification of the species can be performed as NEXAFS spectra are specific for each $TiO_x$ phase.

6 regions of interest (ROIs) were identified, indicated in Fig. 3b and 3c by numbered circles. The Ti $2p$ and O $1s$ X-ray absorption spectra extracted from each of these localized regions are reported in Fig. 3d and 3e, respectively. There are significant variations in the Ti $2p$ spectral shapes clearly indicating significant changes in the chemical structure of $TiO_x$. In particular, peak splitting and changes in relative intensities of the Ti $2p$ peaks can be attributed to changes in symmetry around Ti. These spectral differences were then employed to spatially resolve different $TiO_x$ phases across the $TiO_x$ film. Details of phase identifications at the Ti $2p$ and O $1s$ are reported in Supplementary material. Spectra extracted from undamaged areas of the film far away (ROI_1) and close (ROI_2) to the identified defect, are similar to each other with broad and smooth spectral features indicating the existence of amorphous $TiO_x$.[22,23] The spectra extracted in the defect regions, ROI_3, ROI_4 and ROI_6, are quite different and characteristic of crystalline $TiO_2$. A notable finding is the splitting of the ($2p_{3/2}$, $e_g$) peak, which indicates distortion of the $TiO_6$ octahedra.[26] The relative intensities of these two peaks depend on the particular type of octahedral distortion, therefore on the particular polymorphic form of $TiO_2$ (Supplementary, Fig. S6a).[27] Based on the spectral shape identification, ROI_3 can be assigned to an orthorhombic-like phase such as brookite or $TiO_2$-II, while ROI_4 is assigned to $TiO_2$ rutile and ROI_6 to $TiO_2$ anatase (see Supplementary for details of phase attributions). The spectrum of ROI_5 is quite peculiar and represents the predominant species observed underneath the TE defect. The intensity of ($2p_{3/2}$, $t_{2g}$) peak in the Ti $2p$ spectrum extracted from ROI_5 is lower than that in the spectrum from ROI_3, ROI_4 and ROI_6, indicating lower local symmetry at point 5. The low energy shoulder around 456 eV is consistent with the presence of a partially reduced phase containing $Ti^{3+}$.[13,27] This is also confirmed by the fact that the peaks are slightly broader compared to ROI_3, ROI_4 and ROI_6, as inferred from the reduced $2p_{3/2}$ and $2p_{1/2}$ peak to valley ratio. Spectra of phases containing $Ti^{3+}$ show broader features compared to the pure $Ti^{4+}$ phases because of the binding energies of the $Ti^{3+}$ levels and overlapping and increasing numbers of allowed transitions in the $Ti^{3+}$ spectra.[29,30] Formation of $Ti^{3+}$ species underneath the defect can be explained by the following reaction:

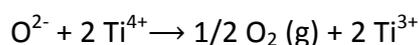
$$O^{2-} + 2\ Ti^{4+} \rightarrow 1/2\ O_2\ (g) + 2\ Ti^{3+}$$

$O_2$ gas accumulates below the TE and erupts from the weakest part of the film, causing the physical dome-like deformation observed at the rim of the TE.[31] The edge of the TE has been suggested recently to be the preferred reduction site since it is a three-phase contact site (oxide, metal and ambient atmosphere).[32] Variations between the selected ROI_s are less evident at the O $1s$ as the change in structure of $TiO_x$ has less influence on O $1s$ than Ti $2p$ spectra (Fig. 3e and S6b).[33] Nevertheless, our analysis of O $1s$ spectra is in good agreement with that of Ti $2p$, as outlined in Supplementary materials.

In order to better visualise the chemical distribution of the main $TiO_x$ phases across the LRS case, the aligned Ti $2p$ and O $1s$ image sequences on an optical density scale were converted to chemical components maps by means of singular value decomposition (SVD) procedures.[34] The black and white chemical maps at the Ti $2p$ (Fig. 4a, b) and O $1s$ (Fig. 4c, d) were obtained using the two predominant components in the film, namely amorphous (corresponding to the spectrum of ROI_1 and ROI_2) and reduced (corresponding to the spectrum of ROI_3) TiO$_x$ phases. The brightest regions correspond to TiO$_x$ amorphous in Fig. 4a, c and TiO$_x$ reduced in Fig. 4b, d. The fit of the two components gives consistent results at the Ti 2$p$ and O 1$s$ edges. In the unaffected regions, TiO$_x$ remains amorphous in between the electrodes whereas the reduced TiO$_x$ phase is localized in a well-defined region underneath the TE protrusion (C) and in the two thinner filamentary regions connecting TE and BE on the left side of the protrusion (A and B).

The spatial correlation of amorphous and reduced TiO$_x$ phases is better visualized by merging the two components in single color-coded composition maps. Spectra of these components at the Ti 2$p$ and O 1$s$ are respectively shown in Fig. 5a and 5b and were used to generate the color-coded composition maps reported in Fig. 5d and 5e, with blue regions corresponding to amorphous and green to reduced TiO$_x$. Localization of the two phases is apparent with the TiO$_x$ amorphous in the unaffected film areas whereas the reduced TiO$_x$ is localized mainly below the highest protrusion of the defect and in the regions A and B. As reduced TiO$_x$ has increased conductivity compared to the amorphous state, this is a direct observation of the area responsible for the LRS state of the device. Our observations are further validated by appending the Ti 2$p$ and O 1$s$ stacks of the same ROI area (Supplementary, Fig. S7) and combining amorphous and reduced TiO$_x$ spectra into a single one (Fig. 5c). The Ti 2$p$ and O 1$s$ shown in Fig. 5f are in excellent agreement giving further evidence of the highly reduced TiO$_x$ percolation.

Further analysis was performed by generating chemical components maps exploring various combinations of reference spectra extracted from the image sequences. For example, the colour-code composite map at the O 1$s$ shown in Supplementary, Fig. S8b, was obtained by including in the fitting the O 1$s$ spectrum of the SiO$_2$ layer below the BE (Fig. S8a). A satisfactory final fit of the Pt(BE)/TiO$_x$/Pt(TE) cross-section was obtained by adding a third component in addition to amorphous and reduced TiO$_x$ (Supplementary, Fig. S9). The Ti 2$p$ spectrum of the third component (represented in red in Fig. S9a and S9b) is similar to that of the reduced phase but exhibits a smaller 2$p_{1/2}$ crystal field splitting and it has a significant lower intensity compared to the reduced TiO$_x$. The O 1$s$ spectrum of the third component is quite different from that of reduced TiO$_x$. Combined color-coded maps obtained independently at the Ti 2$p$ and O 1$s$ (Fig. S9c and S9d), show that the third phase, represented in red, is clearly present only in the TE and BE regions. This also explains the observed lower intensity in comparison to the other two components. Combined fitting at Ti 2$p$ and O 1$s$ further validates this finding (Fig. S9e). Although this clearly demonstrates the presence of Ti in both Pt TE and BE, the identification of this phase is not straightforward. The O 1$s$ spectrum suggests the formation of a new phase, perhaps a Pt alloy.

The combined approach allows us to estimate the O/Ti ratio of the three species used for the fitting by performing an atom-edge jump analysis (Supplementary, Table S1). Our results indicate that the O/Ti ratio is 1.85±0.15 for the amorphous film, 1.37±0.15 for the reduced and 5±1 for the reduced phase in the Pt BE and TE. Therefore, the combined cross chemical analysis is consistent with the elemental compositions of TiO$_{1.85}$ for amorphous TiO$_x$, close to Ti$_2$O$_3$ for the reduced region and TiO$_5$ for the phase in the BE and TE. Chemical mapping results combined with point spectra analysis of selected ROI_s show that the core of the conductive filament is formed by a reduced phase with a O/Ti ratio close to Ti$_2$O$_3$ adjacent to a crystalline phases such as TiO$_2$ rutile and brookite or TiO$_2$-II in the region between the filament and the delaminated TE area. The Ti$_2$O$_3$ phase has been suggested by others as the phase forming the conductive filaments.[35,36,37] It is also interesting to point out that the Magnéli phase, Ti$_4$O$_7$, often considered as the phase forming the conductive filament, is a homogeneous solid solution of (TiO$_2$) rutile and Ti$_2$O$_3$ phases.[38,39] In particular, its structure consists of rutile-like layers of TiO$_6$ octahedra which share faces, edges and corners, similar to the Ti$_2$O$_3$ corundum structure, with the titanium charge nearly equally distributed among the rutile and Ti$_2$O$_3$-like sites.[40] It is also interesting to note that the profile of region ROI_3 is very close to that of orthorhombic TiO$_2$-II which has been recently prepared by ALD [41] and found to be an intermediate phase in rutile formation by ball milling.[42] Its presence could support the argument of local stress manifested due to the switching. The presence of crystalline phases adjacent to the reduced TiO$_x$ area confirms that the current passes through the reduced filament and the heating is mainly localized in this region. The increase in temperature necessary for the growth of crystalline TiO$_2$ phases is induced by the localized Joule heating produced during the switching cycles. It was previously shown that by annealing a TiO$_2$ amorphous film, the anatase phase is formed first because of energy surface minimization around 650 K.[43,44] Formation of rutile only occurs after further annealing at higher temperatures, around 1000 K.[17] In our film, the anatase phase is only observed in a region slightly away from the core of the filament, where a small reduced area is observed. However the rutile phase was identified in the region adjacent to the main reduced filament. This suggests that the Joule heating temperature reached in the filament as a consequence of the bias reaches 1000 K.

By recording a sequence of X-ray images of the cross-section of the TiO$_x$ thin film at closely spaced photon energies, we have directly visualized the location of the reduced nanofilament and characterized it chemically at a very fine spatial scale by extracting Ti 2$p$ (450-485 eV) and O 1$s$ (525-555 eV) X-ray absorption spectra. Our results demonstrate that the conductive filament is composed by a reduced TiO$_x$ phase with a O/Ti ratio close to Ti$_2$O$_3$ and that it is located right underneath a protrusion of the top electrode. Crystalline TiO$_2$ rutile phase and an additional orthorhombic-like TiO$_2$ phase regions were identified for the first time, in a localized area between the TiO$_x$ reduced phase and the top electrode of the device. This provides the first experimental indication that the temperature in the core of the filament increases up to 1000 K. A Ti-containing phase similar to the TiO$_x$ reduced phase was also identified in both the top and bottom Pt electrodes, giving evidence of Ti diffusion into the Pt contact electrodes due to the switching. This technique can be adapted for other RRAM materials, enabling more in-depth studies in reduction dynamics.

**Figure 1** Device architecture and electrical characterisation of solid-state RRAM devices. **(a)** Schematic of the device stack structure. **(b-c)** Optical images of TiO$_x$ RRAM stand-alone devices. **(d-e)** Electrical characterization of devices before TXM measurement: **(d)** pristine and **(e)** LRS state devices.

**Figure 2** Experimental set-up. **(a)** SEM image of pristine device active junction. **(b)** SEM images of final thin lamella. **(c)** Lamella orientation for the TXM-NEXAFS experiment. **(d)** Schematic of the TXM for NEXAFS experiment.

**Figure 3** Physicochemical characterization of a RRAM device programmed in LRS. Optical density TXM micrographs of LRS device obtained with photon energy of 450 eV **(a)** and 458 eV **(b, c)**. NEXAFS Ti 2*p* **(d)** and O 1*s* **(e)** spectra extracted from the regions circled in the X-ray images b and c. Scale bar 150 nm **(a, b)**, 100 nm **(c)**.

**Figure 4** Chemical maps depicting regions of amorphous and reduced TiO$_x$. Components maps at the Ti 2*p* **(a, b)** and at the O 1*s* **(c, d)**. Scale bar = 100 nm.

**Figure 5** Physicochemical characterization of a RRAM device programmed in LRS. Color-coded composition maps at the Ti 2*p* and O 1*s* generated independently **(d, e)** and simultaneously **(f)** by using the absorption spectra of the two components, TiO$_x$ amorphous (A) and TiO$_x$ reduced (B) shown in **(a), (b), (c)**. Blue (amorphous TiO$_x$) and green (reduced TiO$_x$). Scale bar = 100 nm.

## References

1. Yang, J. J., Strukov, D. B. & Stewart, D. R. Memristive devices for computing. *Nat. Nanotech.* **8,** 13–24 (2013).
2. Strukov, D. B., Snider, G. S., Stewart, D. R. & Williams, R. S. The missing memristor found. *Nature* **453,** 80–83 (2008).
3. Chanthbouala, A. *et al.* A ferroelectric memristor. *Nat. Mater.* **11,** 860–864 (2012).
4. Jo, S. H. *et al.* Nanoscale memristor device as synapse in neuromorphic systems. *Nano Lett.* **10,** 1297–1301 (2010).
5. Prodromakis, T., Toumazou, C. & Chua, L. Two centuries of memristors. *Nat. Mater.* **11,** 478–481 (2012).
6. Wong, H.-S. P. & Salahuddin, S. Memory leads the way to better computing. *Nat. Nanotechnol.* **10,** 191–194 (2015).
7. Yang, J. J. *et al.* Memristive switching mechanism for metal/oxide/metal nanodevices. *Nat. Nanotechnol.* **3,** 429–433 (2008).
8. Lee, M.-J. *et al.* A fast, high-endurance and scalable non-volatile memory device made from asymmetric Ta$_2$O$_{5-x}$/TaO$_{2-x}$ bilayer structures. *Nat. Mater.* **10,** 625–630 (2011).
9. Zhao, L. *et al.* Multi-level control of conductive nano-filament evolution in HfO$_2$ ReRAM by pulse-train operations. *Nanoscale* **6,** 5698–5702 (2014).
10. Jeong, D. S. *et al.* Emerging memories: resistive switching mechanisms and current status. *Rep. Prog. Phys.* **75,** 076502 (2012).

11. Schroeder, H., Pandian, R. & Miao, J. Resistive switching and changes in microstructure. *Phys. Status Solidi A* **208,** 300–316 (2011).
12. Kwon, D. *et al.* Atomic structure of conducting nanofilaments in $TiO_2$ resistive switching memory. *Nat. Nanotechnol.* **5,** 148–153 (2010).
13. Strachan, J. P. *et al.* Direct identification of the conducting channels in a functioning memristive device. *Adv. Mater.* **22,** 3573–3577 (2010).
14. Strachan, J. P. *et al.* Characterization of electroforming-free titanium dioxide memristors. *Beilstein J. Nanotechnol.* **4,** 467–473 (2013).
15. Koehl, a. *et al.* Evidence for multifilamentary valence changes in resistive switching $SrTiO_3$ devices detected by transmission X-ray microscopy. *APL Mater.* **1,** 042102 (2013).
16. Hitchcock, A. P., Dynes, J. J., Johansson, G., Wang, J. & Botton, G. Comparison of NEXAFS microscopy and TEM-EELS for studies of soft matter. *Micron* **39,** 741–748 (2008).
17. Kavei, G., Nakaruk, A. & Sorrell, C. C. Equilibrium State of Anatase to Rutile Transformation for Titanium Dioxide Film Prepared by Ultrasonic Spray Pyrolysis Technique. *Mater. Sci. Appl.* **2,** 700–705 (2011).
18. Jo, Y. *et al.* Resistance switching mode transformation in $SrRuO_3$/Cr-doped $SrZrO_3$/Pt frameworks via thermally activated Ti out-diffusion process. *Sci. Rep.* **4:7354,** 1–7 (2014).
19. Hitchcock, A. P. in *Soft X-ray Imaging Spectromicroscopy Handb. Nanoscopy* (ed. Gustaaf Van Tendeloo, D. V. D. and S. J. P.) 745–791 (Wiley, 2012).
20. Ade, H. & Hitchcock, A. P. NEXAFS microscopy and resonant scattering : Composition and orientation probed in real and reciprocal space. *Polymer (Guildf).* **49,** 643–675 (2008).
21. Groot, F. M. F. De, Fuggle, J. C., Thole, B. T. & Sawatzky, G. A. $L_{2,3}$ x-ray-absorption edges of $d^0$ compounds: $K^+$, $Ca^{2+}$, $Sc^{3+}$, and $Ti^{4+}$ in $O_h$ (octahedral) symmetry. *Phys. Rev. B* **41,** 928–937 (1990).
22. Kucheyev, S. *et al.* Electronic structure of titania aerogels from soft x-ray absorption spectroscopy. *Phys. Rev. B* **69,** 245102 (2004).
23. Wang, D., Liu, L. & Sham, T. Observation of lithiation-induced structural variations in $TiO_2$ nanotube arrays by X-ray absorption fine structure. *J. Mater. Chem. A* **3,** 412–419 (2015).
24. Waser, R. & Aono, M. Nanoionics-based resistive switching memories. *Nat. Mater.* **6,** 833–840 (2007).
25. Kim, K. M., Jeong, D. S. & Hwang, C. S. Nanofilamentary resistive switching in binary oxide system; a review on the present status and outlook. *Nanotechnology* **22,** 254002 (2011).
26. Crocombette, J. P. & Jollet, F. Ti 2p x-ray absorption in titanium dioxides ($TiO_2$): The influence of the cation site environment. *J. Phys. Condens. Matter* **6,** 10811–10821 (1994).
27. Groot, D. 2p X-ray Absorption of Titanium in Minerals. *Phys. Chem. Miner.* **19,** 140–147 (1992).
28. Strachan, J. P. *et al.* The switching location of a bipolar memristor: chemical, thermal and structural mapping. *Nanotechnology* **22,** 254015 (2011).
29. Abbate, M. *et al.* Soft-x-ray-absorption studies of the location of extra charges induced by substitution in controlled-valence materials. *Phys. Rev. B* **44,** 5419–5422 (1991).
30. De Groot F.M.F.; Grioni M.; Fuggle J.C. Oxygen 1s x-ray-absorption. *Phys. Rev. B* **40,** 5715–5723 (1989).
31. Joshua Yang, J. *et al.* The mechanism of electroforming of metal oxide memristive switches. *Nanotechnology* **20,** 215201 (2009).
32. Lenser, C. *et al.* Insights into nanoscale electrochemical reduction in a memristive oxide: The role of three-phase boundaries. *Adv. Funct. Mater.* **24,** 4466–4472 (2014).

33. De Groot F.M.F., Faber J., Michielis J.J.M., Czyzyk M.T., Abbate M., F. J. C. Oxygen 1s x-ray absorption of tetravalent titanium oxides: A comparison with single-particle calculations. *Phys. Rev. B* **48,** 2074–2080 (1993).
34. Koprinarov, I. N., Hitchcock, A. P., Mccrory, C. T. & Childs, R. F. Quantitative Mapping of Structured Polymeric Systems Using Singular Value Decomposition Analysis of Soft X-ray Images. *J. Phys. Chem. B* **106,** 5358–5364 (2002).
35. Mazady, A. & Anwar, M. Memristor: Part I-The Underlying Physics and Conduction Mechanism. *IEEE Trans. Electron Devices* **61,** 1054–1061 (2014).
36. Zhong, X., Rungger, I., Zapol, P. & Heinonen, O. Electronic and magnetic properties of $Ti_4O_7$ predicted by self-interaction-corrected density functional theory. *Phys. Rev. B* **115143,** 1–8 (2015).
37. Szot, K. *et al.* $TiO_2$ -a prototypical memristive material. *Nanotechnology* **22,** 254001 (2011).
38. Lucovsky, G., Miotti, L. & Bastos, K. P. Detection of multivalency charge states in complex and elemental transition metal oxides by X-ray absorption spectroscopy: Controlled multivalency as a pathway to device functionality. *Jpn. J. Appl. Phys.* **50,** 10PF04 (2011).
39. Liborio, L., Mallia, G. & Harrison, N. Electronic structure of the $Ti_4O_7$ Magnéli phase. *Phys. Rev. B* **79,** 245133 (2009).
40. Marezio, M., Dernier, P. D., Laboratories, B. T., Hill, M. & Remeika, J. P. The Crystal Structure of $Ti_4O_7$ a Member of the Homologous Series $Ti_nO_{2n-1}$. *J. Solid State Chem.* **3,** 340–348 (1971).
41. Aarik, J., Aidla, A., Uustare, T. Atomic-layer growth of $TiO_2$ -II thin films. *Philos. Mag. Lett.* **73,** 115–119 (1996).
42. S. Begin-Colin Girot, T., Mocellin, A. & Caer, G. Le. Kinetics of formation of nanocrystalline $TiO_2$ II by high energy ball-milling of anatase $TiO_2$. *Nanostructured Mater.* **12,** 195–198 (1999).
43. Ranade, M. R. *et al.* Energetics of nanocrystalline $TiO_2$. *PNAS Colloq.* **99,** 6476–6481 (2001).
44. Rath, S. *et al.* Angle dependence of the O K edge absorption spectra of $TiO_2$ thin films with preferential texture. *Nucl. Instruments Methods Phys. Res. Sect. B* **200,** 248–254 (2003).
45. Guttmann, P. *et al.* Nanoscale spectroscopy with polarized X-rays by NEXAFS-TXM. *Nat. Photonics* **6,** 25–29 (2012).
46. Jacobsen, C., Wirick, S., Flynn, G. & Zimba, C. Soft X-ray spectroscopy from image sequences with sub-100 nm spatial resolution. *J. Microsc.* **197,** 173–184 (2000).
47. Hitchcock, A. aXis2000. *Available Free noncommercial use from http//unicorn.mcmaster.ca/aXis2000.html* at <http://unicorn.mcmaster.ca/aXis2000.html>

## Methods

### Device Preparation

The Cr/Pt/TiO$_x$/Pt based memristor devices were fabricated on an oxidised (200 nm SiO$_2$) 6-inch Si wafer. The bottom electrode (BE) and top electrode (TE) were fabricated using conventional optical lithography, electron-beam evaporation followed by a lift-off process, with BE and TE composed of Cr/Pt (3 nm/30 nm) and Pt (30 nm) successively. The Cr film in the BE served as adhesive layer for Pt. A 50 nm TiO$_x$ layer was then deposited using reactive sputtering from a Ti metal target with the following settings: 8 sccm O$_2$, 35 sccm Ar, 2 kW at the cathode, and 15 sccm O$_2$ and 2 kW at the additional plasma source.

### Electrical switching of devices

Electrical biasing of the devices has been carried out through voltage sweeping in order to assess or modify their resistive states as required (2 V maximum). Passive resistive state assessment of the Device Under Test (DUT) was carried out via low voltage, non-invasive voltage sweeps whilst resistive switching was induced through the application of more aggressive, high voltage sweeps. All voltage biasing was carried out under current compliance protection. The compliance current level was determined on *ad hoc* basis. All sweeping was performed using a Keithley 4200 electrical characterisation instrument.

### FIB-SEM

A dual-beam focused ion beam / scanning electron microscope system (Zeiss NVision 40 FIB/FEGSEM) equipped with a gas injection system (GIS) was used to record SEM images and for cutting FIB cross-sections. SEM images were recorded at an accelerating voltage of 5 kV. Prior to performing FIB cross-sections, an electron beam-induced tungsten protective layer was deposited on the top of the electrodes in order to minimize damage caused by the gallium ions in the subsequent ion beam-induced tungsten deposition step. After extraction, the thickness of the lamella is further decreased to allow X-ray transmission by low energy ion polishing at a low incident angle until a thickness of 40-70 nm is achieved.

### TXM-NEXAFS

The TXM-NEXAFS study was performed at the undulator beamline U41-FSGM at the BESSY II electron storage ring operated by the Helmholtz-Zentrum Berlin, Germany. The optical design of TXM has a spatial resolution of 25 nm and spectral resolution E/ΔE = 10000.[45] The Ti 2$p$ images were acquired from 450 to 485 eV in 0.1 eV steps of photon energy. In order to increase the signal to noise ratio of Ti 2$p$ spectra, three separate images were acquired for each photon energy giving a total of 1053 images. The O 1$s$ images were acquired from 525 to 540 eV in 0.1 eV steps and from 540 to 555 eV in 0.2 eV steps giving a total of 351 images. Images size is 1340x1300 pixels size corresponding to 5 nm per pixel. Each image was taken with the sample at the proper focus position to optimize spatial resolution.[46] First, a stack of all images was created followed by automatic alignment using Fourier cross correlation techniques in aXis2000 and in Stack_Analyze. Alignment of images is required to correct for lateral motion of the X-ray beam on the sample.[46] The I$_o$ for each stack was obtained from an internal region of each stack (off the FIB section) and used to convert the aligned stack from transmission to optical density (OD = -log[I(E)/Io(E)]). Ti 2$p$ and O 1$s$ spectra were then extracted from the stack. Spectra can be obtained from regions as small as the spatial resolution of the microscope (25 nm). Data analysis was performed using software aXis2000.[47]

**AFM**

AFM maps were acquired with a MultiMode Nanoscope V AFM (Veeco Metrology Group) in contact mode using Pt/Ir coated Si tips with a cantilever spring constant of 0.2 N/m and nominal radius of 12 nm (Bruker, SCM-PIC).

## Acknowledgments

The authors wish to thank the HZB, Germany for the allocation of synchrotron radiation beamtime as well as Dr. S. Werner and Dr. K. Henzler, both HZB, for their support during the beamtime. We also wish to acknowledge the support of Dr. S. Boden, Southampton Nanofabrication Centre, University of Southampton, UK and Dr. J. C. Walker, Faculty of Engineering and the Environment, University of Southampton, UK for their support during lamellae preparation. We finally wish to acknowledge the financial support of the EPSRC EP/K017829/1 and EU-FP7 RAMP.

## Author contributions

D.C. and T.P. conceived the experiment. D.C. performed the synchrotron experiment and analysed the data. A.P.H. extended the aXis2000 program to facilitate this study and performed data analysis, in particular chemical mapping. P.G. gave support during synchrotron data acquisition and discussion of the analysis. A.R. gave experimental assistance in performing the synchrotron experiment and analysed XPS data. A.K. fabricated the devices. A.S. and I.G. performed electrical characterisation of devices. D.C. and T.P. wrote the manuscript.

## Competing financial interests

The authors declare no competing financial interests.

**Figure 1**

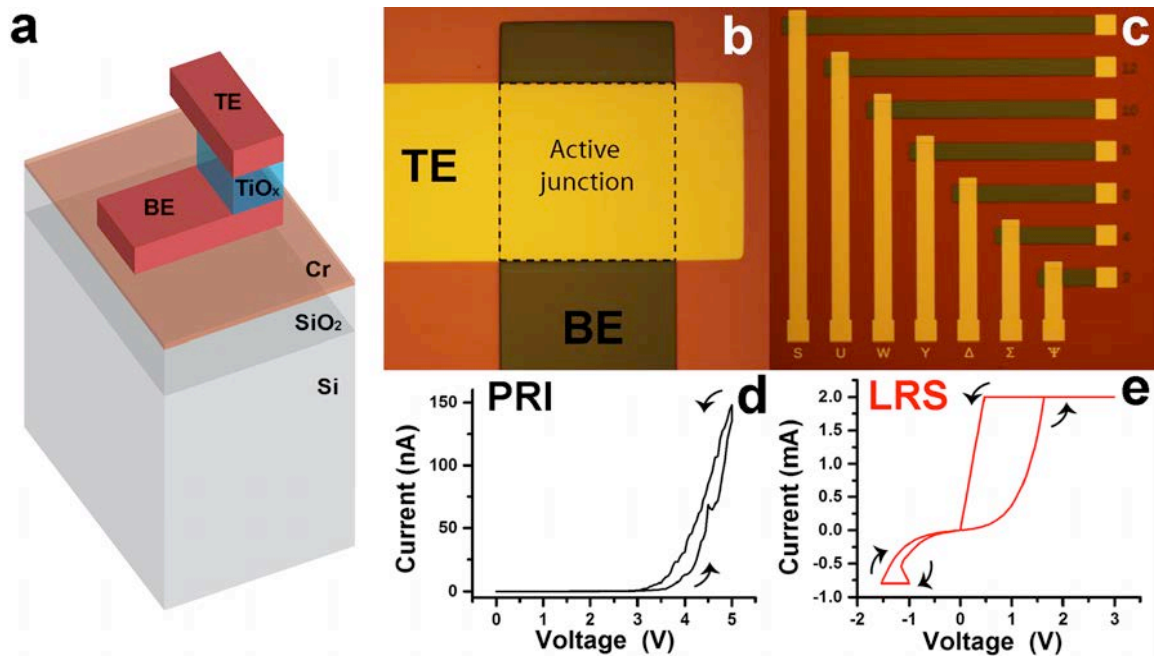

**Figure 2**

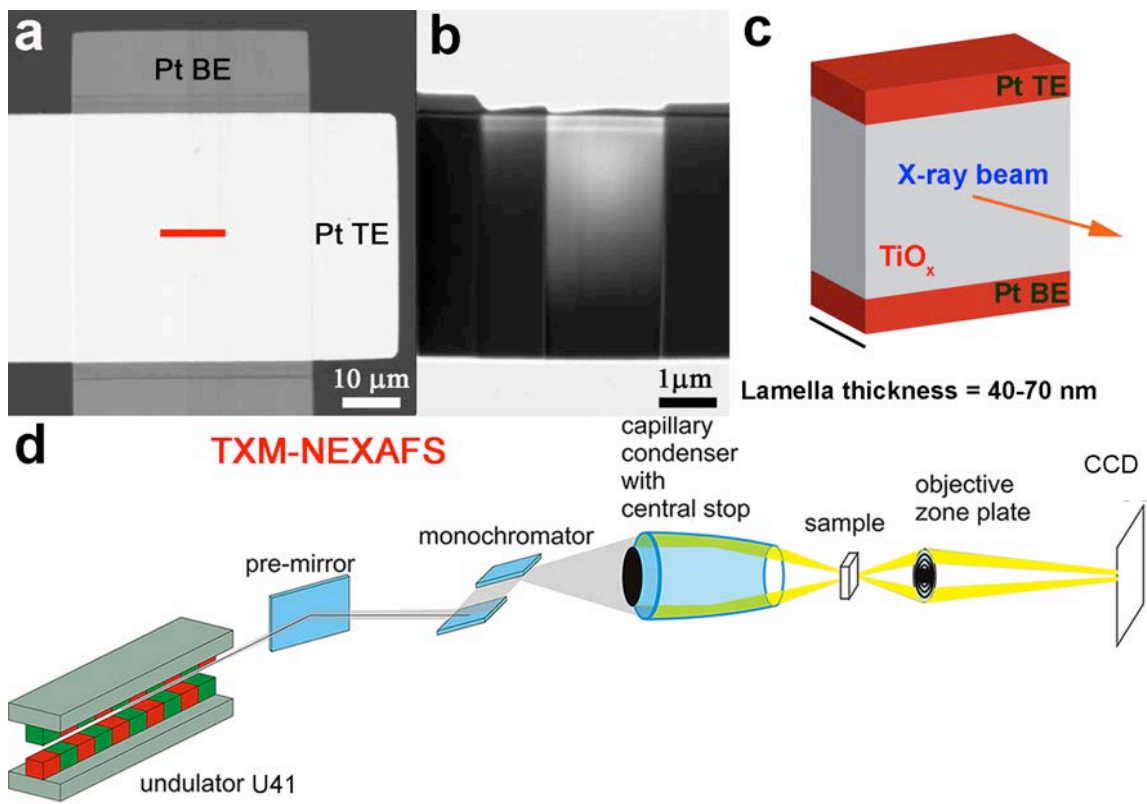

**Figure 3**

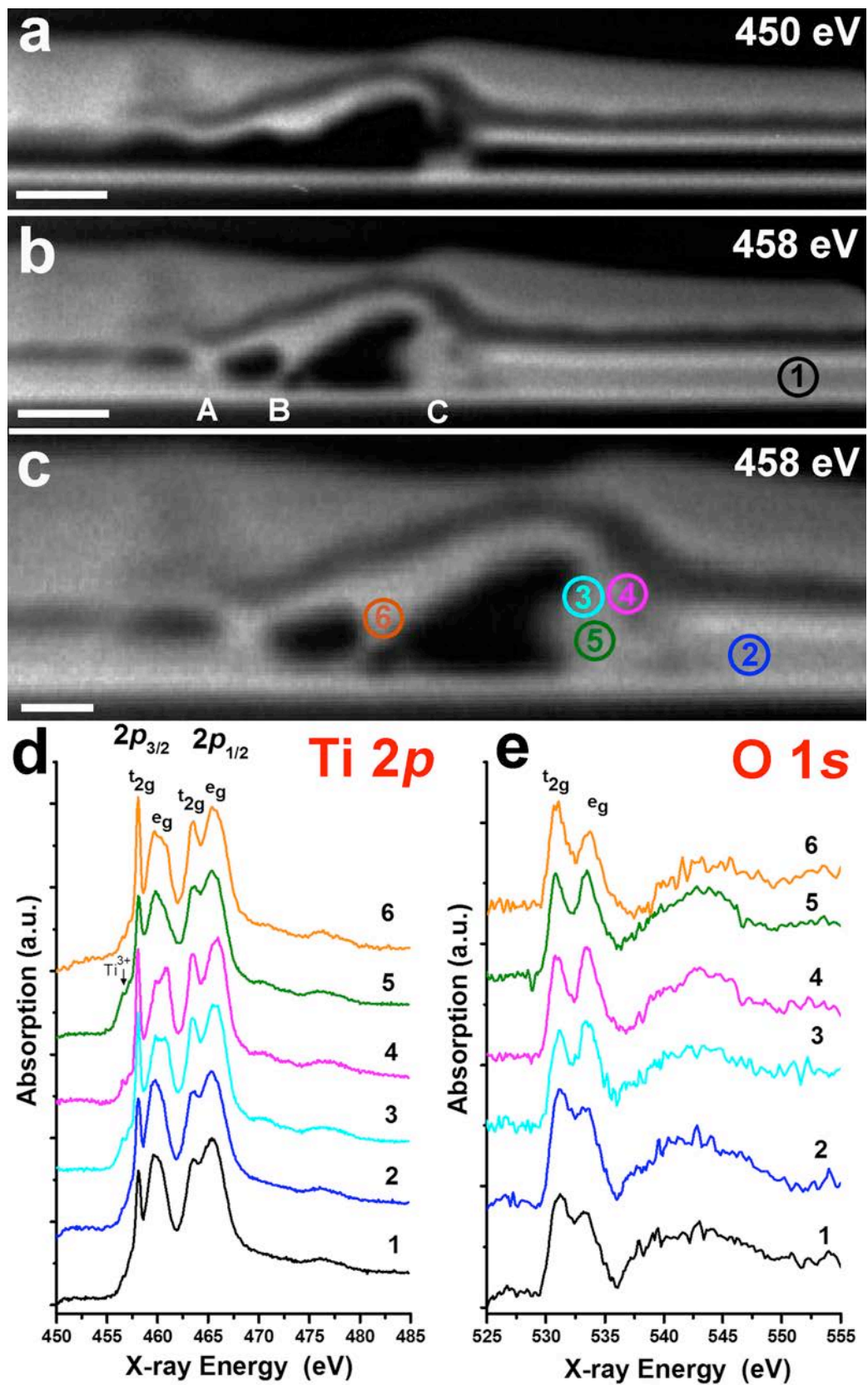

**Figure 4**

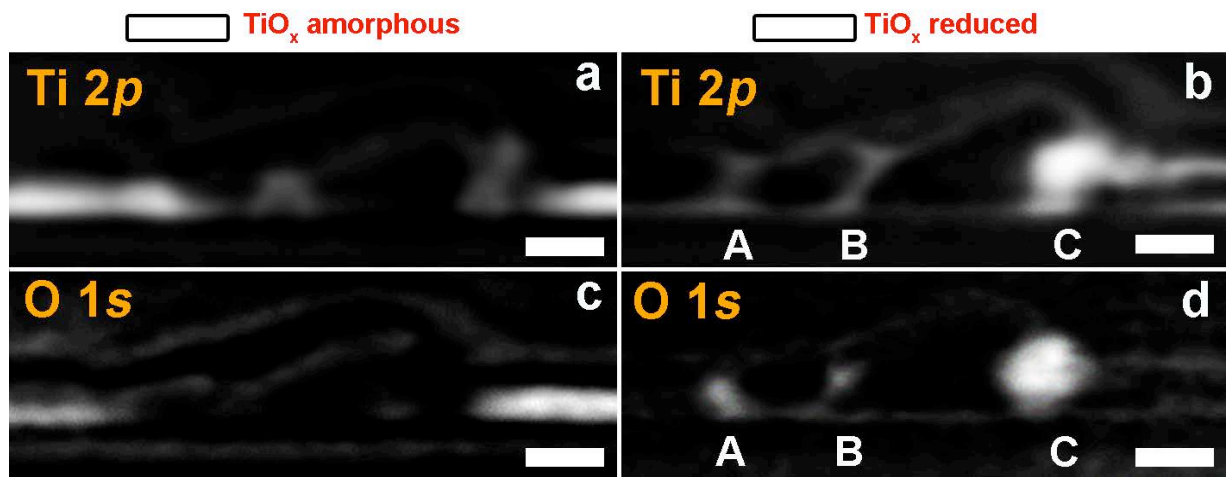

**Figure 5**

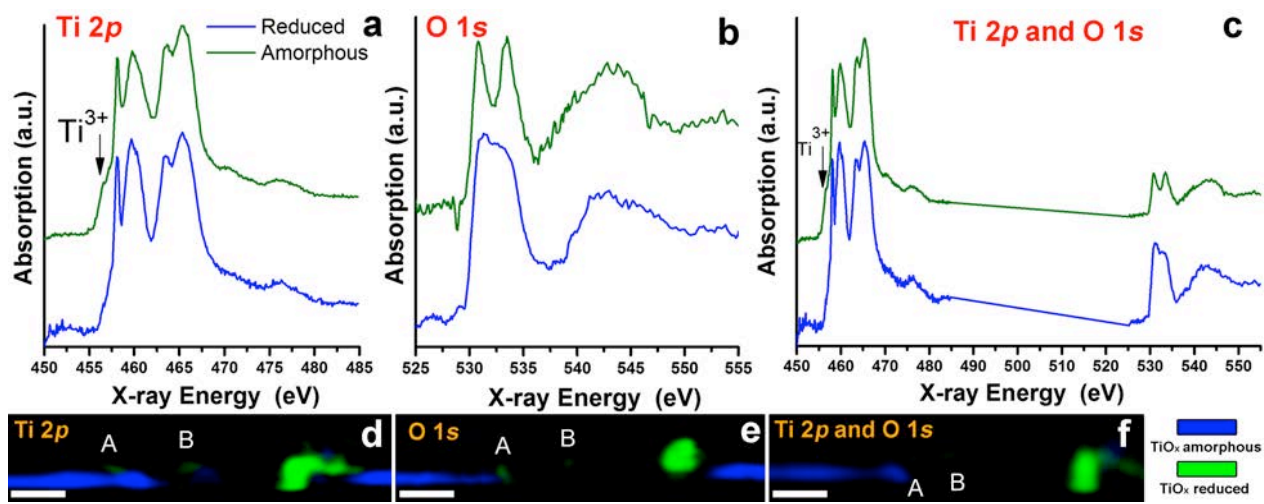

## Supplementary Information

## Spatially resolved TiO$_x$ phases in RRAM conductive nanofilaments using soft-X-ray spectromicroscopy

Daniela Carta, Adam Hitchcock, Peter Guttmann, Anna Regoutz, Ali Khiat, Alexantru Serb, Isha Gupta, Themistoklis Prodromakis

1. **X-ray photoelectron spectroscopy (XPS)**

XPS was used to characterise the TiO$_x$ thin film. All spectra were recorded on a Thermo Scientific Theta Probe Angle-Resolved X-ray Photoelectron Spectrometer (ARXPS) system with a monochromated Al Kα X-ray source (hv = 1486.6 eV). The X-ray source was operated at 6.7 mA emission current and 15 kV anode bias and pass energies of 200 eV and 50 eV were used for survey and core level spectra, respectively. Spectra were corrected for any charge shifts by aligning them to the C 1$s$ core level at 285.0 eV and all data were analysed using the Avantage software package. The XPS survey spectrum of the TiO$_x$ thin film along with the O 1$s$ and Ti 2$p$ core levels are shown in Fig. S1a, S1b and S1c, respectively. The Ti 2$p$ core level shows charge transfer satellites S$_{3/2}$ and S$_{1/2}$ at higher binding energies.[1] Furthermore, a small population of Ti$^{3+}$ of the order of 4 % of the total Ti is observed (Fig. S1d). The O 1$s$ core level shows a surface oxygen component, often referred to as non-lattice oxygen, on the higher binding energy side of the main core line.[2]

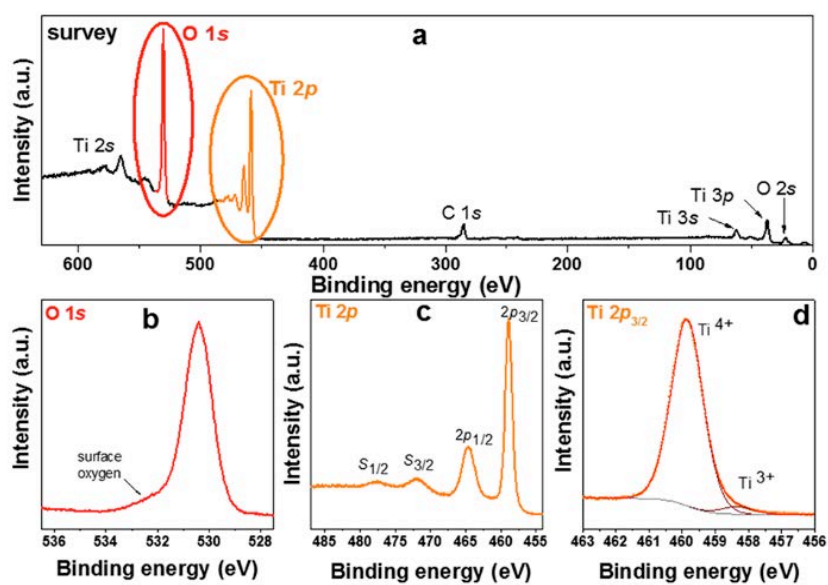

**Figure S1** XPS spectra of TiO$_x$. **(a)** Survey spectrum of TiO$_x$ showing all core levels. **(b)** O 1$s$ core level. **(c)** Ti 2$p$ core level including higher binding energy satellites S$_{3/2}$ and S$_{1/2}$. **(d)** Fit of the Ti 2$p_{3/2}$ peak.

## 2. TXM-NEXAFS

The TXM set-up used in this work presents significant advantages compared with previously used geometries. In particular, it does not require replacement of the silicon wafer support with a fragile standing $Si_3N_4$ window [3,4] (which could affect the device electrical behaviour due to strain effects and difficulty in sinking the Joule heating [5,6]) and does not require removal of the top electrode prior to analysis (which is usually performed by a scotch tape method and could lead to unwanted removal of the thin $TiO_2$ layer underneath critical areas [5,7]), both necessary steps to enable the X-ray transmission if irradiating the device from the top electrode. Most importantly, our geometry allows direct visualization and chemical investigation of the cross-section of the device, as shown in Fig. S2a.

Ti 2$p$ and O 1$s$ spectra extracted from the $TiO_x$ film in PRI device case are shown in Fig. S2c and S2d, respectively. The first doublet ($2p_{3/2}$) (457-462 eV) of Ti 2$p$ spectra (Fig. S2c) originates from transitions to ($2p_{3/2}$, 3$d$-$t_{2g}$) and ($2p_{3/2}$,3$d$-$e_g$) states while the second doublet ($2p_{1/2}$) (462-468 eV) originates from transitions to the corresponding $2p_{1/2}$ states. The $2p_{3/2}$ – $2p_{1/2}$ splitting is due to spin-orbit coupling while the $t_{2g}$-$e_g$ separation is the crystal-field splitting due to the surrounding O atoms. [8,9] It has to be noted that in all spectra, the ($2p_{3/2}$, $e_g$) peak is broader than the ($2p_{3/2}$, $t_{2g}$) due to the large degree of hybridization of $e_g$ orbitals with O ligand orbitals. [10] Satellite peaks at 470.5 and 476.0 eV due to polaronic transitions are also often observed. [11,12] The O 1$s$ spectra (Fig. S2d) can be divided in two regions. The double between 528 and 536 eV can be attributed to O 1$s$ excitation to hybrid excited states in which the final level is a mixture of O 2$p$ and Ti 3$d$ orbitals. The spectral features at 531.3 and 533.4 eV are assigned to the $t_{2g}$ and $e_g$ orbitals, respectively. [13] This region is very sensitive to local symmetry and coordination. Peaks in the region between 536 eV and 555 eV corresponds to O 1$s$ excited states in which the final level is a hybridization of O 2$p$ and Ti 4$sp$ orbitals.[14] This region is more sensitive to long-range order.[13]

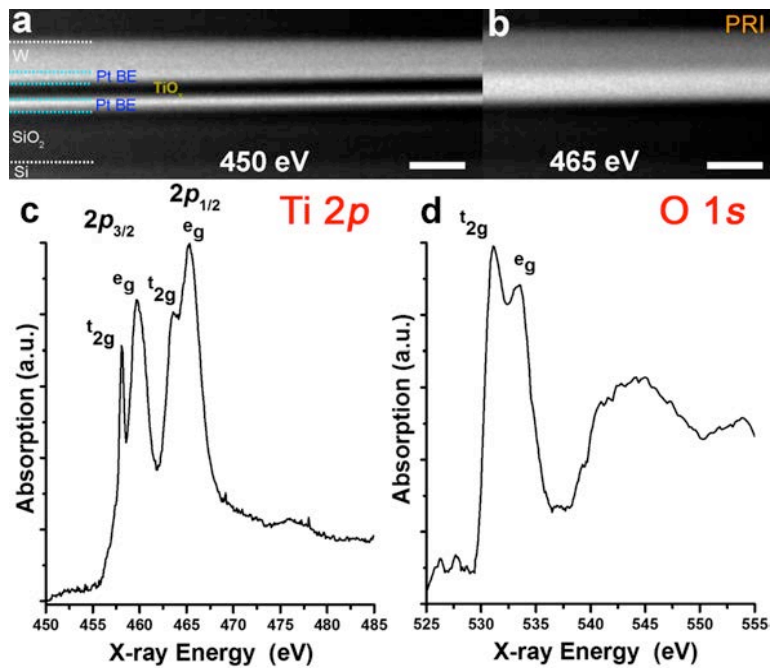

**Figure S2** Optical density TXM micrographs of the PRI case obtained using X-ray photon energy of 450 eV **(a)** and 465 eV **(b)** and Ti 2$p$ **(c)** and O 1$s$ **(d)** NEXAFS spectra extracted from the $TiO_x$ film. Scale bar = 200 nm.

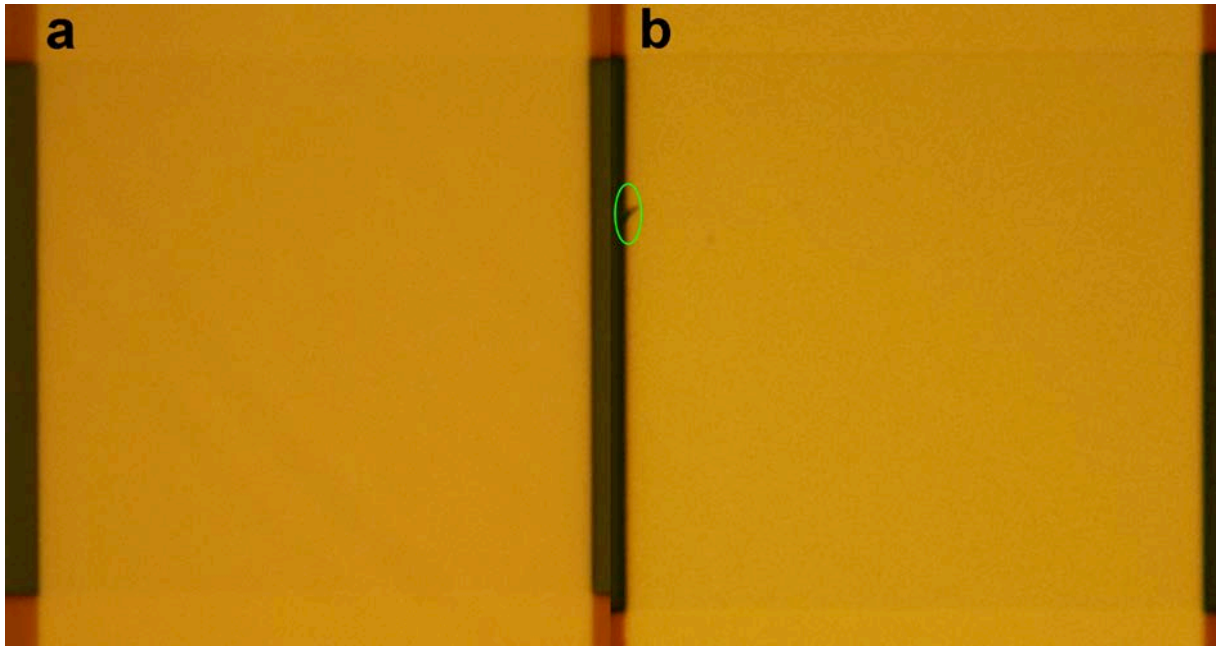

**Figure S3** Optical images of LRS device before **(a)** and after **(b)** switching into LRS. Morphological defect of TE is circled in green.

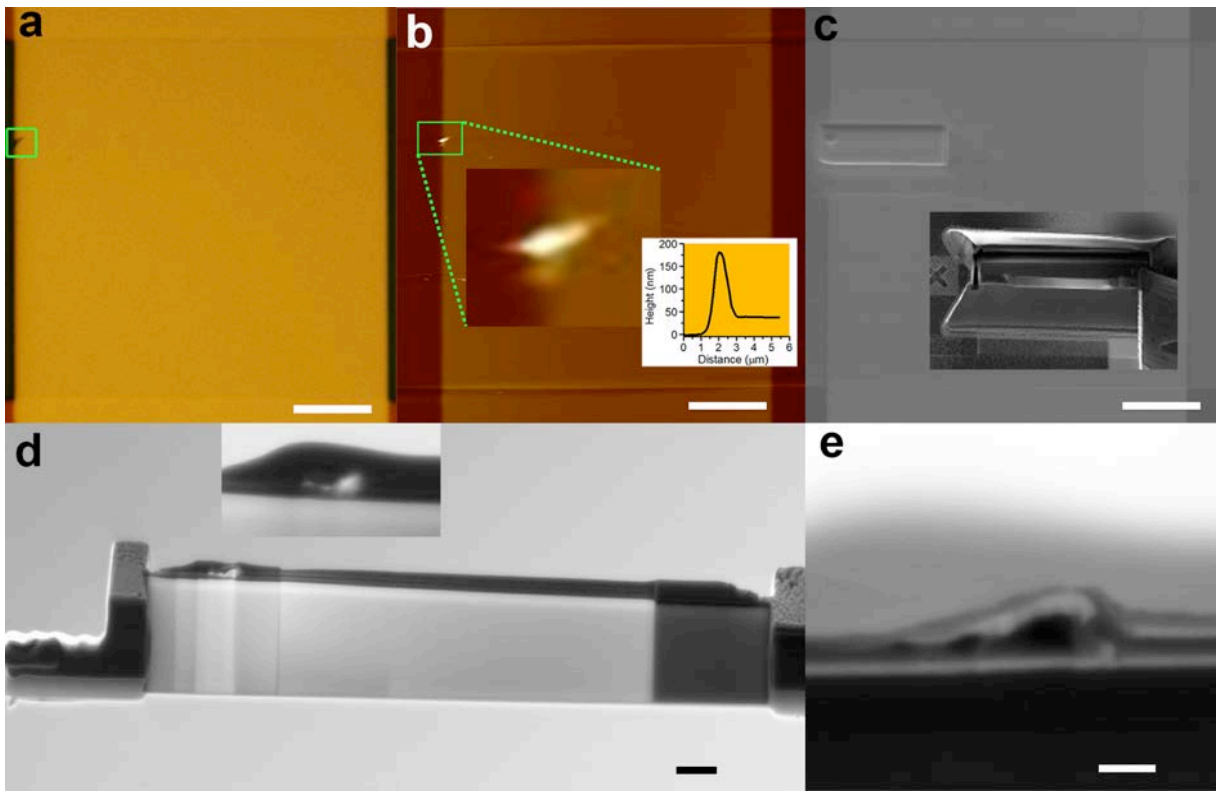

**Figure S4 (a)** Optical and **(b)** AFM images viewed from the TE. Inset in **b**: line profile of the protrusion. **(c)** SEM image of lamella cut location and in inset detail of lamella extraction. **(d)** Thinned lamella with details of the damaged region. **(e)** Cross-section SEM image of the damaged region. Scale bar **(a-c)** 10 μm; **(d)** 1 μm; **(e)** 200 nm.

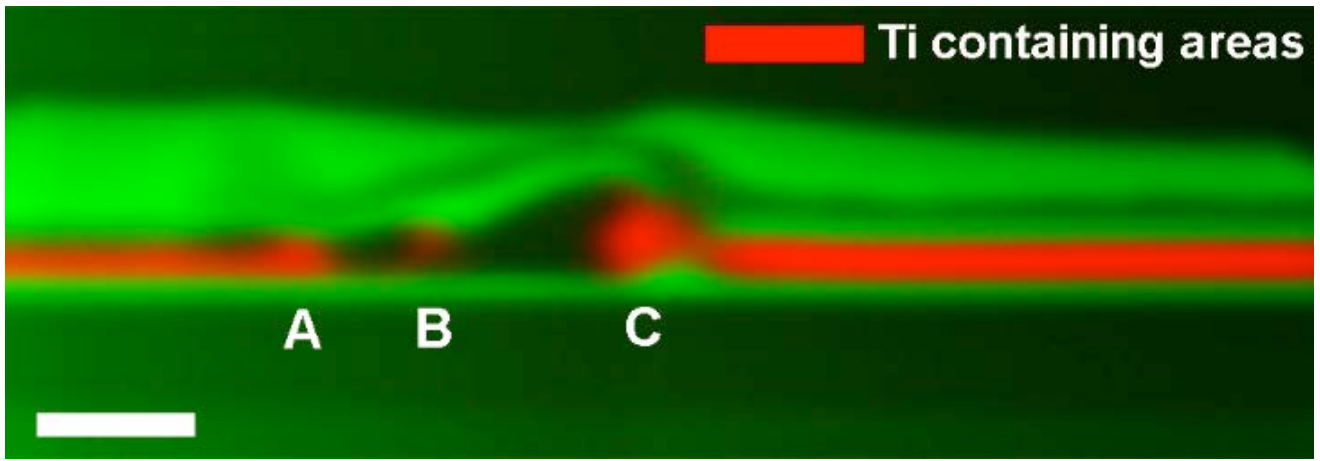

**Figure S5** Average of all images of the Ti stack showing in red the regions containing only Ti species. Scale bar = 200 nm.

## 3. Spectral changes of TiO$_2$ polymorphs

**Ti 2p**

Peak splitting and change in relative intensities of the Ti 2p peaks can be attributed to change in symmetry around Ti. In particular, the (2$p_{3/2}$, $e_g$) orbitals at the Ti 2p point directly towards the 2p orbitals of the surrounding O and are therefore very sensitive to the titanium local environment (Fig. S6a).[15,10] In the spectra of ROI_2 (and the similar ROI_1), the higher intensity of the (2$p_{3/2}$, $e_g$) peak compared to that of the (2$p_{3/2}$, $t_{2g}$) peak [16] and the absence of splitting of the (2$p_{3/2}$, $e_g$) peak around 460 eV indicate a high structural disorder as expected for amorphous TiO$_x$. Splitting of the (2$p_{3/2}$, $e_g$) peak, observed in ROI_3, ROI_4 and ROI_6 indicates octahedron distortion of the oxygen ligands in TiO$_6$.[17] The relative intensities of these two peaks depend on the particular type of octahedral distortion. In the spectrum of ROI_3, the intensities of these two peaks are comparable. This particular (2$p_{3/2}$, $e_g$) profile has been reported for orthorhombic-like phases such as TiO$_2$-II [18] and brookite.[19] The profile of ROI_3 region is very similar to that of the TiO$_2$-II phase reported in [18], an usually high pressure phase which has recently been grown at ambient pressure by atomic layer deposition as 50 nm thin films.[20] In the spectrum of ROI_4, the relative intensities of the split (2$p_{3/2}^{-1}$, $e_g$) peak are different, being higher for the peak at higher energy. This is typical of TiO$_2$ rutile phase which has tetragonal distortion of the TiO$_6$ octahedra.[10,19,21] Assignment of this spectrum to the rutile phase is also supported by the fact that the crystal-field splitting of the 2$p_{1/2}$ peak is 2.4 eV.[17] Finally, in the spectrum of ROI_6, the relative intensity of the split (2$p_{3/2}$, $e_g$) peak is opposite to that in rutile, being the intensity higher for the peak at lower energy. This profile is typical of anatase. Moreover, the crystal-field splitting of the 2$p_{1/2}$ peak is 1.9 eV, lower than the one observed in rutile, and typical of the anatase phase.[18,17] Two additional low intensity peaks appear in ROI_3, ROI_4 and ROI_6 at 456.6 and 457.2 eV (indicated as a and b in Fig. S6a). They have been observed previously in rutile, anatase [10] and TiO$_2$-II [18] and their origin is not fully understood. It has been suggested that they could be due to strong interaction between 3d electrons and the 2p core hole [22] or to sharp features in the 3d partial density of states.[23]

## O 1s

The corresponding O 1s X-ray absorption spectra extracted from the same localized regions used to extract the Ti 2p edge spectra are reported in Fig. S6b. O 1s spectra of undamaged areas ROI_2 (and the similar ROI_1) show broad and smooth spectral features typical of amorphous $TiO_x$.[24] Moreover, the dip between the (O 1s, $t_{2g}$) and (O 1s, $e_g$) peaks is shallow compared to the spectra of regions from ROI_3 to ROI_6 and their energy difference, which is often used to evaluate crystal field splitting, is smaller and typical of disordered materials (2.0 eV).[15] In the spectra extracted from ROI_3 to ROI_6, the $e_g$ and $t_{2g}$ peaks appear more defined and intense. The energy gap between $e_g$ and $t_{2g}$ in ROI_4 and ROI_6 is ~ 3.0 eV, a value reported for both anatase and rutile phases.[10] This gap is slightly smaller in ROI_3.

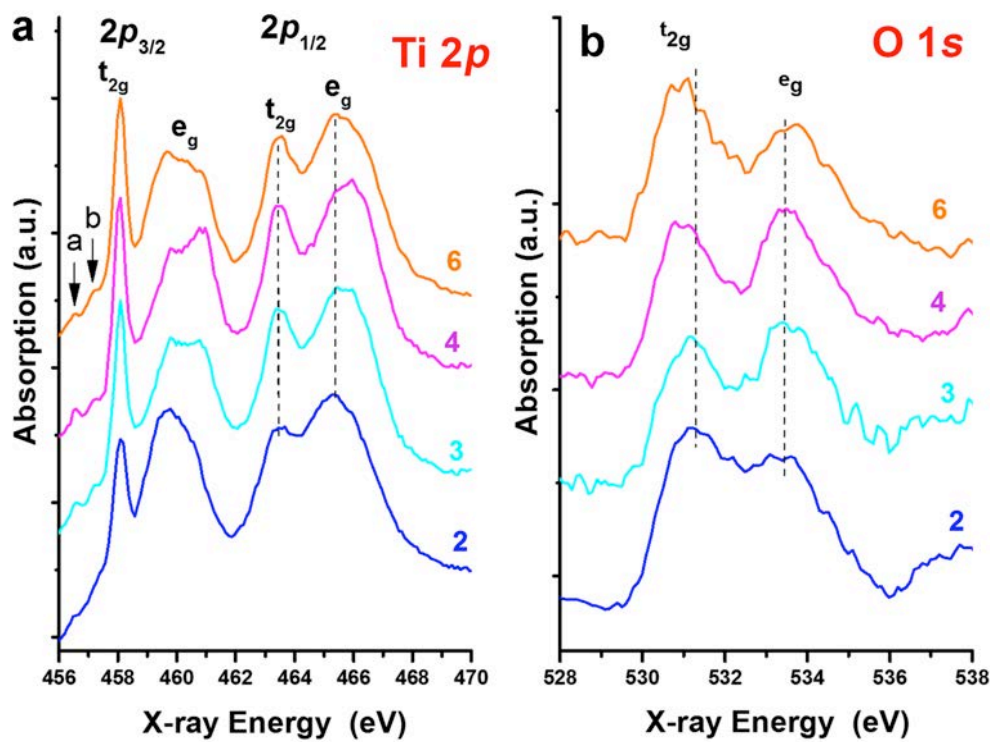

**Figure S6 |** NEXAFS point Ti 2p (a) and O 1s (b) spectra extracted from the ROI_2, ROI_3, ROI_4 and ROI_6 circled in the X-ray image of Fig. 3c in the main manuscript.

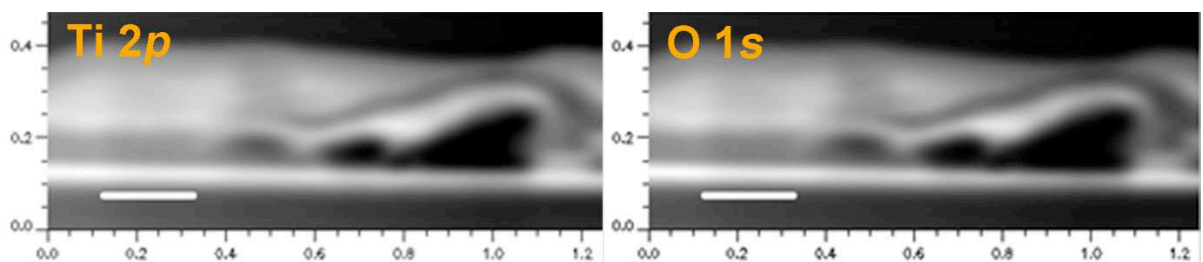

**Figure S7 |** Ti 2p and O 1s stacks adjusted to the same spatial area and mesh size corresponding to 4.75 nm pixel size, used for combined stack of Ti 2p and O 1s. The final size for both stack is 291x93 pixels. Scale bar = 200 nm.

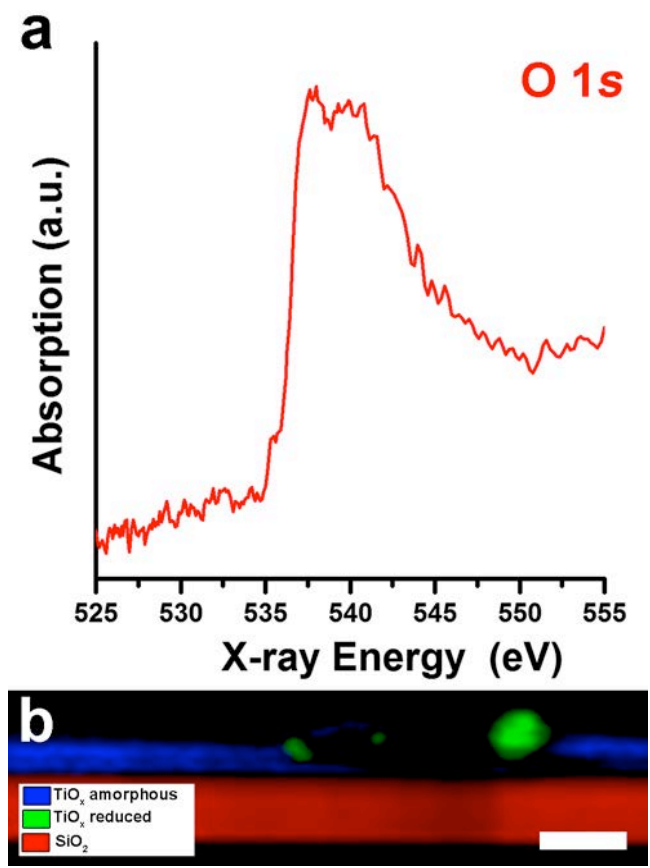

**Figure S8 | (a)** O 1*s* spectrum from the SiO$_2$ layer support. **(b)** Color-coded composition map of selected components: red (SiO$_2$), green (reduced TiO$_x$) and blue (amorphous TiO$_x$). Scale bar = 100 nm.

### 4. Chemical mapping

The best fit at the Ti 2*p* and O 1*s* was obtained using three components, TiO$_x$ amorphous, TiO$_x$ reduced and a third component similar to TiO$_x$ reduced but slightly different at the O 1*s* (Fig. S9). If additional TiO$_x$ components were added, the component maps from the fit contained large regions with unphysical negative coefficients, indicative of an over-determined fit. If any one of the three reference spectra were removed from the fit, the residual of the fit increased significantly in the regions of the missing component.

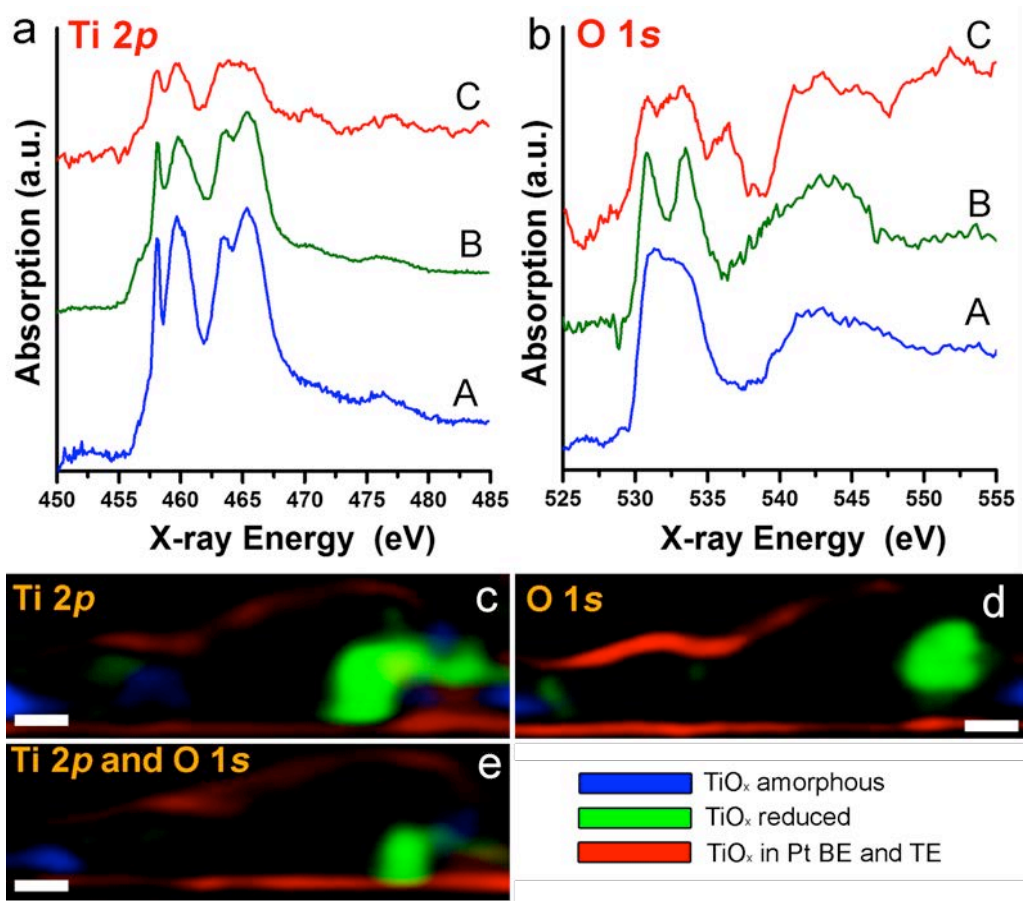

**Figure S9 |** NEXAFS Ti 2*p* **(a)** and O 1*s* **(b)** spectra of amorphous TiO$_x$ (A), reduced TiO$_x$ (B) and the Ti containing phase observed in the TE and BE (C). **(c-e)** Color-coded composition maps of selected components at the Ti 2*p* and O 1*s* generated independently **(c, d)** and combining Ti 2*p* and O 1*s* **(e)**. Blue (amorphous TiO$_x$), green (reduced TiO$_x$) and red (phase in Pt electrodes). Scale bar = 50 nm.

|   | Amorphous TiO$_x$ | | | Reduced TiO$_x$ | | | Phase in TE/BE | | |
|---|---|---|---|---|---|---|---|---|---|
|   | Jump | Thickness (nm) | O/Ti | Jump | Thickness (nm) | O/Ti | Jump | Thickness (nm) | O/Ti |
| Ti | 0.17 | 74±5 |  | 0.09 | 40±5 |  | 0.06 | 24±4 |  |
| O | 0.26 | 137±8 | 1.85±0.15 | 0.10 | 55±5 | 1.37±0.15 | 0.22 | 116±6 | 5±1 |

**Table S1.** Edge jumps, absolute thickness of OD maps and O/Ti nominal ratio of amorphous and reduced TiO$_x$ and third phase in TE and BE. Error were estimated from reproducibility with small variations in reference spectra.

# References

1. Okada, K., Uozumi, T. & Kotani, A. Split-Off State Formation in the Final State of Photoemission in Ti Compounds. *J. Phys. Soc. Japan* **63,** 3176–3184 (1994).

2. Park, S.-J. *et al.* In situ control of oxygen vacancies in $TiO_2$ by atomic layer deposition for resistive switching devices. *Nanotechnology* **24,** 295202 (2013).

3. Strachan, J. P. *et al.* Direct identification of the conducting channels in a functioning memristive device. *Adv. Mater.* **22,** 3573–3577 (2010).

4. Strachan, J. P. *et al.* Characterization of electroforming-free titanium dioxide memristors. *Beilstein J. Nanotechnol.* **4,** 467–473 (2013).

5. Strachan, J. P. *et al.* Structural and chemical characterization of $TiO_2$ memristive devices by spatially-resolved NEXAFS. *Nanotechnology* **20,** 485701 (2009).

6. Borghetti, J. *et al.* Electrical transport and thermometry of electroformed titanium dioxide memristive switches. *J. Appl. Phys.* **106,** 124504 (2009).

7. Koehl, a. *et al.* Evidence for multifilamentary valence changes in resistive switching $SrTiO_3$ devices detected by transmission X-ray microscopy. *APL Mater.* **1,** 042102 (2013).

8. Sánchez-Santolino, G. *et al.* Characterization of surface metallic states in SrTiO3 by means of aberration corrected electron microscopy. *Ultramicroscopy* **127,** 109–113 (2013).

9. Stoyanov, E., Langenhorst, F. & Steinle-Neumann, G. The effect of valence state and site geometry on Ti $L_{3,2}$ and O K electron energy-loss spectra of $Ti_xO_y$ phases. *Am. Mineral.* **92,** 577–586 (2007).

10. Kucheyev, S. *et al.* Electronic structure of titania aerogels from soft x-ray absorption spectroscopy. *Phys. Rev. B* **69,** 245102 (2004).

11. Das, C., Tallarida, M., Schmeißer, D. Linear dichroism in ALD layers of $TiO_2$. *Environ. Earth Sci.* **70,** 3785–3795 (2013).

12. Laan, G. Van Der. Polaronic satellites in x-ray-absorption spectra. *Phys. Rev. B* **41,** 12366–12368 (1990).

13. Lusvardi, V. S. *et al.* An NEXAFS investigation of the reduction and reoxidation of $TiO_2$(001). *Surf. Sci.* **397,** 237–250 (1998).

14. Chen, X. *et al.* Properties of disorder-engineered black titanium dioxide nanoparticles through hydrogenation. *Sci. Rep.* **3,** 1510 (2013).

15. Fusi, M. *et al.* Surface electronic and structural properties of nanostructured titanium oxide grown by pulsed laser deposition. *Surf. Sci.* **605,** 333–340 (2011).

16. Groot, F. M. F. De, Fuggle, J. C., Thole, B. T. & Sawatzky, G. A. 2p x-ray absorption of 3d transition-metal compounds: An atomic multiplet description including the crystal field. *Phys. Rev. B* **42,** (1990).

17. Crocombette, J. P. & Jollet, F. Ti 2p x-ray absorption in titanium dioxides ($TiO_2$): The influence of the cation site environment. *J. Phys. Condens. Matter* **6,** 10811–10821 (1994).

18. Ruus R. Kikas A., Saar A., Ausmees A., Nommiste. E, Aarik A., Aidla A., Uustare T., M. I. Ti 2p and O 1s absorption of $TiO_2$ polymorphs. *Solid State Commun.* **104,** 199–203 (1997).

19. Henderson, G. S., Liu, X. & Fleet, M. E. A Ti L-edge X-ray absorption study of Ti-silicate glasses. *Phys. Chem. Miner.* **29,** 32–42 (2002).

20. Tarre, A. *et al.* Atomic layer deposition of epitaxial $TiO_2$ II on c-sapphire. *J. Vac. Sci. Technol. A* **31,** 01A118 (2013).

21. Lopez, M. F. *et al.* Soft x-ray absorption spectroscopy study of oxide layers on titanium alloys. *Surf. Interface Anal.* **33,** 570–576 (2002).

22. Groot, F. M. F. De, Fuggle, J. C., Thole, B. T. & Sawatzky, G. A. $L_{2,3}$ x-ray-absorption edges of $d^0$ compounds: $K^+$, $Ca^{2+}$, $Sc^{3+}$, and $Ti^{4+}$ in $O_h$ (octahedral) symmetry. *Phys. Rev. B* **41,** 928–937 (1990).

23. Finkelstein, L. D. *et al.* Vacant states of $TiO_2$ with rutile structure and their reflection in different-type x-ray absorption spectra. *X-Ray Spectrom.* **31,** 414–418 (2002).

24. Wang, D., Liu, L. & Sham, T. Observation of lithiation-induced structural variations in $TiO_2$ nanotube arrays by X-ray absorption fine structure. *J. Mater. Chem. A* **3,** 412–419 (2015).